\documentclass[11pt,aps,pre,preprint]{revtex4-1}
\usepackage{amsmath} 
\usepackage[dvips]{graphicx}
\usepackage{natbib}
\usepackage{color}

\begin{document}
\title{\sf Transforming complex network to the acyclic one}

\author{Roman Shevchuk$^{1,2}$ }

\author{Andrew Snarskii$^2$ \footnote{Corresponding authors; E-mail: asnarskii@gmail.com, roman.shevchuk@frias.uni-freiburg.de;Phone: +49 (0)761 203 97336, Fax:
+49 (0)761-203 97451}}
\affiliation{$^{1}$ Freiburg Institute for Advanced Studies,School of Soft Matter Research, Albertstrasse 19, 79104 Freiburg im Breisgau,
Germany}
\affiliation{$^{2}$ Kyiv Politechnical Institute, Prospect Peremogy 37, 03056, Kyiv, Ukraine.}
\date{\today}

\title{\sf Transforming complex network to the acyclic one}
\begin{abstract}
Acyclic networks are a class of complex networks in which  links are directed and don't have closed loops. Here we present an algorithm for transforming an ordinary undirected complex network  into an acyclic one. Further analysis of an acyclic network allows finding structural properties of the network.  With our approach one can find the communities and key nodes in complex networks. Also we propose a new parameter of complex networks which  can mark most vulnerable nodes of the system. The proposed algorithm can be applied to finding communities and bottlenecks in general complex networks.

\end{abstract}
\date{\today}

\maketitle

\section{Introduction}
Acyclic networks are a class of complex networks in which all the links are directed and don't have closed loops. As an example the network of food chains is the acyclic one: edges are directed from anode representing a predator to the node representing its prey \cite{Milo2002Network}. Another example of an acyclic complex network is the network of citation: this is a directed network in which every edge has a direction \cite{Lehmann2003Citation}.

Basic research in the field of acyclic complex network is focused on networks with a hierarchical structure. Systems, organization of which displays causal asymmetry constraints can be described in terms of directed links using a discrete set of arbitrary units, the evolutionary history of a given collection of events or the chart  of computational states visited along a complex simulation. Such a set of paths and links defines the acyclic network \cite{CRGS10,Costa2007Characterization,KN09}. Acyclic complex networks are widely used for different physical problems: from protein folding to sociophysics \cite{Karrer2009Random, Krivov2006OneDimensional}. One of the fundamental problems of the complex networks is how to divide network into its constituent communities. In the complex networks communities are the groups of nodes that are densely connected amongst themselves while being sparsely connected to the rest of the network. There are different methods to detect communities based on calculation of betweenness, loops of vertexes, analysis of adjacency matrices of the network \cite{girvan:community-structure-social-bio-networks,Radicchi2004Defining,SR95,Newman_PRE74,GfellerD_MDnetworks_2007,Newman2011,Fortunato2010rev,Schaeffer2007, Prada-Gracia}.

In this paper we propose a method to map a general complex network into acyclic graph, arranging them in strong hierarchical order. Mapping a complex networks into the acyclic one opens new opportunities to analyze them which were not present in the usual complex networks. The calculation of these new parameters allows more detailed analysis of the structural characteristics of a complex network. With the proposed transformation of the network it's possible to detect most vulnerable nodes of the structure and detect cluster composition of the network. With our approach one can find the communities, hubs and key nodes in complex networks. Also we propose a new parameter of complex networks which can mark most vulnerable nodes of the system. The proposed algorithm gives a clear picture of an investigated systems, describing some structure details of the complex network.

\section{Methods}
An acyclic network is a graph in which the links have direction and do not form loops. In order to transform a complex network into acyclic one, it is proposed to apply to two nodes of the network the potential difference that will flow from its current. The direction of current flow through the links will set the direction of the links, so in this way the edges don't form a loop. If the network nodes are not connected with the others, then no current will flow through them and acyclicity will not be broken.

To apply current we choose two nodes which we call input and output and put their Voltage to 1 and 0 respectively. The resistance of each link is inverse to its weight:
\begin{equation}
R_{ij}=\frac{1}{a_{ij}}
\end{equation}
Then the Kirchhoff equation can be written as:
\begin{equation}
V_{input}=1
\end{equation}
\begin{equation}
V_{output}=0
\end{equation}
\begin{equation}
V_{i}=\frac{1}{k_{i}}\sum_{i,j \in E}V_{j} =   \frac{1}{k_{i}} \sum_{j \in G} V_{j}a_{ij},
\end{equation}
where $k_{i}$ is node degree and $a_{ij}$ is the adjacency matrix of the network. In general solving the Kirchhoff equations takes $O(n^{3})$, but there was proposed the method when the voltage values can be obtained in linear time \cite{Wu2004Finding}. After calculating all the potentials we arranged all the nodes in the network in the acyclic sequence according to their potential (See Fig.1 for example). To characterize the whole network we chose two nodes, which were the most important for our study. As an example they can be two servers of computer network, if we are interested in the channel capacity between them, or they can represent folded and unfolded states of protein in conformation space network \cite{Rao2004Protein}.

After applying the potential difference we obtained the sequence of nodes and we could relabel them  according to their voltage. Once the nodes are ranked, we can work with transformed network. In \cite{Karrer2009Random} there was introduced the notion of setting the node flux $ \lambda_{i}$ and the flux between two selected nodes $\mu_{i}$, which is defined as:

\begin{equation}
\lambda_i=\sum \limits_{j=1}^{i-1}k_{j}^{in}-\sum \limits_{j=1}^{i}k_{j}^{out}
\end{equation}

\begin{equation}
\mu_i=\sum \limits_{j=1}^{i-1}k_{j}^{in}-\sum \limits_{j=1}^{i-1}k_{j}^{out},
\end{equation}

where $k^{in}$ and $k^{out}$ stand for the links going out  from the node and going in respectively and $k_{j}$ is the index of the node in acyclic sequense i.e. node ranking according to its potential. 
Parameter $\lambda_{i}$ characterizes the amount of links that connect the nodes to the i-th and after it.(See Fig.2 for visual explanation)

\section{Results}
For analysis we created Watts-Strogatz and Erdos-Renyi networks with number of nodes from 200 to 500 and the following probabilities of connection 1\% for Watts-Strogatz and 1\%,2\%,3\% for Erdos-Renyi networks \cite{ER59,Watts1999Small}. To facilitate calculation in our model, all nonzero edges have resistance $R=1$. For a single run (potential difference was applied only to one pair of nodes) we observe different behavior for Watts-Strogatz and Erdos-Renyi (Fig.3). In Watts-Strogatz networks the number of links is smaller comparing to Erdos-Renyi and additional links which were randomly put in the network  create small separated clusters that can be observed. This effect becomes stronger with the increasing of the probability of adding random links, which is crucial for creating subclusters in this type of networks.

For Erdos-Renyi type of networks a single-well structure was observed and we can conclude that the connectivity is homogeneous and there are no well defined clusters. The analysis show no principal difference for all possible pairs in Erdos-Renyi networks. On the other hand, small-world networks' node flux demonstrates splitting the network into many small clusters. So we can make a conclusion that with the observed node flux behavior it's possible to detect different clusters in the network. Although, if both key nodes are taken from the same cluster – one will get single-well picture, so single-run calculation is not enough to describe the whole structure of the system.

For the further observation a  network consisting of two Erdos-Renyi subnetworks was created, which connected a small number of links (comparing to the quantity of links in the subnetworks). Calculated node flux (Fig. 4) accurately describes the structure of the created system and size of its components.  We  analyzed this network to check how sensible was the obtained structure with different pairs of input and output nodes. For the first calculation the nodes belonging to the different subnetworks were taken as input and output  and then we took pairs of nodes by random. In both cases we took 1000 pairs of nodes. Averaged results for both measurements are shown in Fig.5. For pairs of nodes from different subnetworks it is easy to observe the structure of the network. For the other case picture is blurred, because if both input and output were taken from the same cluster they don't give the picture of the real structure. This problem, however, can be solved by using sufficient statistical data. Of course, the precision and accuracy of the obtained structure in case of many pairs measurement are strongly dependent on particular network. Thus, this approach allows detecting the cluster structure of the network and determine belonging of certain nodes to  cluster in the network.

 It is clear that in case of many communities the general picture of $\lambda_{i}$ will not show the real structure of the network. E.g. if there are two ``parallel communities`` with the respect to the input community, their nodes will be mixed. But in this case our approach can define the community of input node with the same quality as in the case of two communities. So one should perform the algorithm for the i node, get the list $C_i$ of the nodes belonging to the same community and than run algorithm with another input not belonging to the $C_i$.

To test our approach we performed the set of benchmarks proposed by Fortunato et al. \cite{BENCH}. One of the main parameters in this benchmark is the mixing parameter,$ \mu $, which is defined as fraction of the links which node shares with another communities(i.e. if $\mu$=0.8 it means that 80\% of the links of the node go inside its community and 20$\%$ are shared with other communities). Degree distribution of the node and community distribution are taken from the power law distribution with exponents $\gamma$ and $\beta$ respectively .It's easy to see that $\mu=0.5$ or higher our approach will fail because there is no boundary between community of the node and the rest of the network. We tested our approach on the networks with the number of nodes N=1000,$\beta=1,2$ and $\gamma=2,3$(Fig. 7). The algorithm shows good results for values of $\mu \leq 0.3$ but for further values its precision  decreases dramatically. The  results are slightly clearer if the average degree of the node is increased because the boundaries between the communities become more pronounced. The comparison of most popular algorithms was done by Lancichinetti and Fortunato \cite{Fortunato_compar}. Although the results of some popular algorithms \cite{girvan:community-structure-social-bio-networks,Blondel,Radicchi2004Defining,Palla,Donetti,Clauset,Amaral, Rosvall2007,Rosvall2008,Ronhovde} are better, the method proposed here can be performed in linear time which is its main advantage.

We also introduced the “vulnerability” of the single node $\Lambda_i$, that can be used to identify the most vulnerable nodes of the network. Parameter  characterizes the ratio of flux which is flowing through the given node (blue edges) and its total flux (blue and red edges). $\Lambda_{i}$ is defined as the ratio of links that go across the given node and its total degree:

\begin{equation}
\Lambda_i=\frac{\mu_i-\lambda_i+\mu_{i+1}-\lambda_i}{\lambda_i}=\frac{k_i}{\lambda_i}
\end{equation}

  In these terms the proposed parameter can describe the “importance” of the chosen node by indicating the value of the flux which is going from one part of the system to another precisely through the given node. Apparently  if the node is a bottleneck between two or more clusters its vulnerability will be maximum, because all the flow between two parts of the system will be flowing only through the link which is connecting two boundary nodes. To test our hypothesis we calculated the vulnerability for two different networks: normal Erdos-Renyi network and network which has two Erdos-Renyi networks connected by small number of links. In the first case  $\Lambda_{i}$ is maximum at the beginning and at the end of the acyclic sequence, which can be explained by the small number of links connected to the input and output nodes; in the case of two clusters we observed the same regions of high $\Lambda_{i}$ values but also another maximum region in the place of connection of two clusters, which means that the nodes in this region are highly loaded and among important for connection of input and output nodes.

\section{Conclusion}
In this  paper we present a method to transform general complex network into the acyclic one. The proposed method can be used to determine cluster structure of the network and the size of its components. According to our measurements the node-flux shows different trend for Erdos-Renyi and Watts-Strogatz types of complex network in single-run calculations, so we can conclude that it's possible to determine the type of the network. We performed the set of the benchmarks which shows that our method can be used to detect communities in linear time.  The proposed parameter vulnerability can be used to detect key-nodes of the network, especially bottleneck-type networks, not only so-called hubs, which can be detected by measuring average node flux and vulnerability.

\section{Acknowledgment}
We want to thank Diego Prada-Gracia and Francesco Rao for help in  preparing of the manuscript and helpful discussion.

\pagebreak

\maketitle
\section{Figures}

\begin{figure}
\begin{center}
\includegraphics[width=130mm]{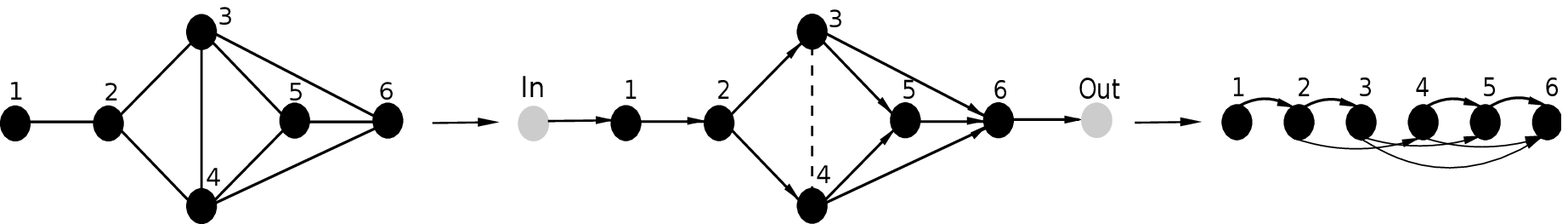}
\caption{Scheme of the transforming of a complex network into an acyclic one.The link $a
_{34}$ was removed because the voltages of the nodes are equal. }
\end{center}
\end{figure}

\begin{figure}
\begin{center}
\includegraphics[width=75mm]{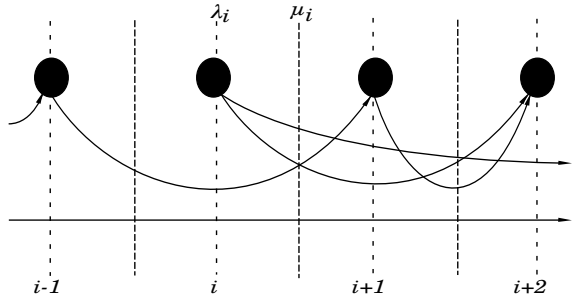}
\caption{Visual representation of the node flux, and the flux between the nodes of a complex network.}
\end{center}
\end{figure}

\begin{figure}[ht!]
\begin{center}
\includegraphics[angle=0,width=1\linewidth]{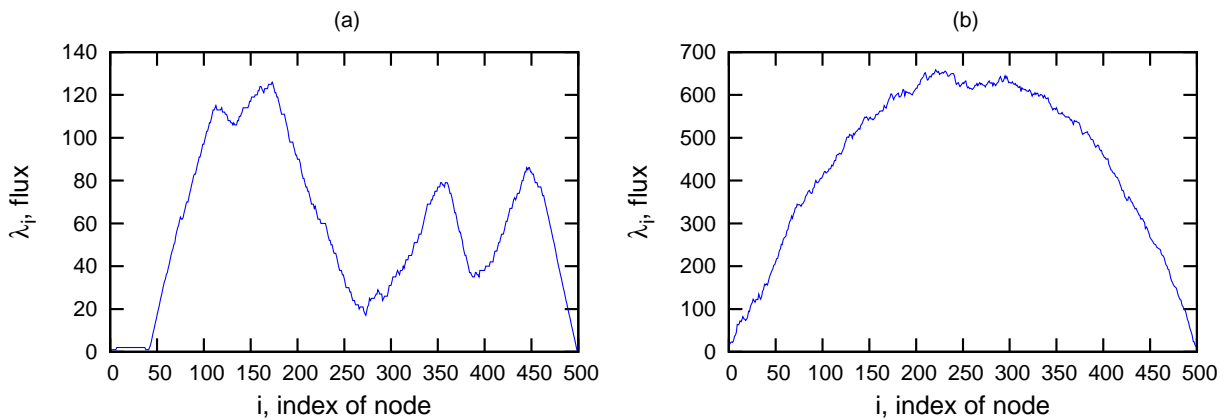}
\caption{Typical plots of  Watts-Strogatz (a) and Erdos-Renyi (b) networks flux, where  $i$ is index of node in the acyclic sequence. }
\end{center}
\end{figure}

\begin{figure}
\begin{center}
\includegraphics[angle=0,width=0.5\linewidth]{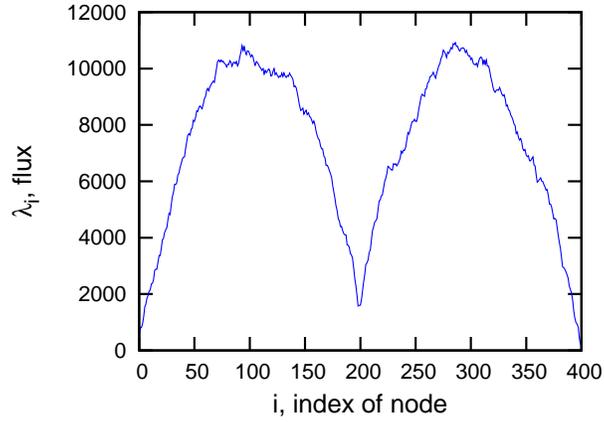}
\caption{Flux of the network, which has two Erdos-Renyi subnetworks connected by small number of links. Each subnetwork is represented by bell-like flux which is typical for Erdos-Renyi networks.$i$ is index of node in the acyclic sequence.}
\end{center}
\end{figure}

\begin{figure}[ht!]
\begin{center}
\includegraphics[angle=0,width=1\linewidth]{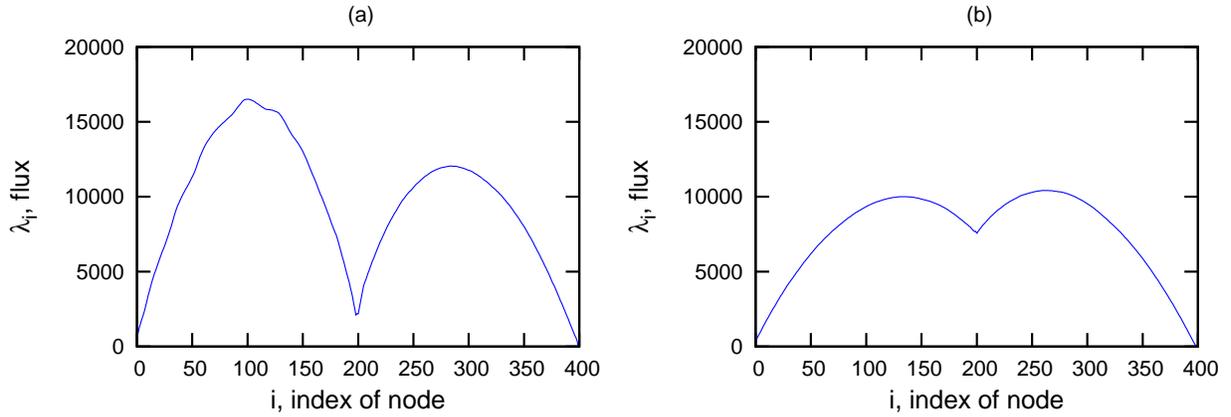}
\caption{Cut with input and output in two different clusters (a) and with randomly chosen input and output (b).}
\end{center}
\end{figure}

\begin{figure}
\begin{center}
\includegraphics[width=75mm]{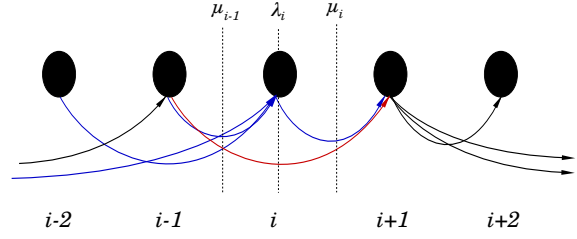}
\caption{Vulnerability  characterizes the ratio of flux which is flowing through the given node (blue edges) and total flux (blue and red edges).}
\end{center}
\end{figure}

\begin{figure}[ht!]
\begin{center}
\includegraphics[angle=270,width=1\linewidth]{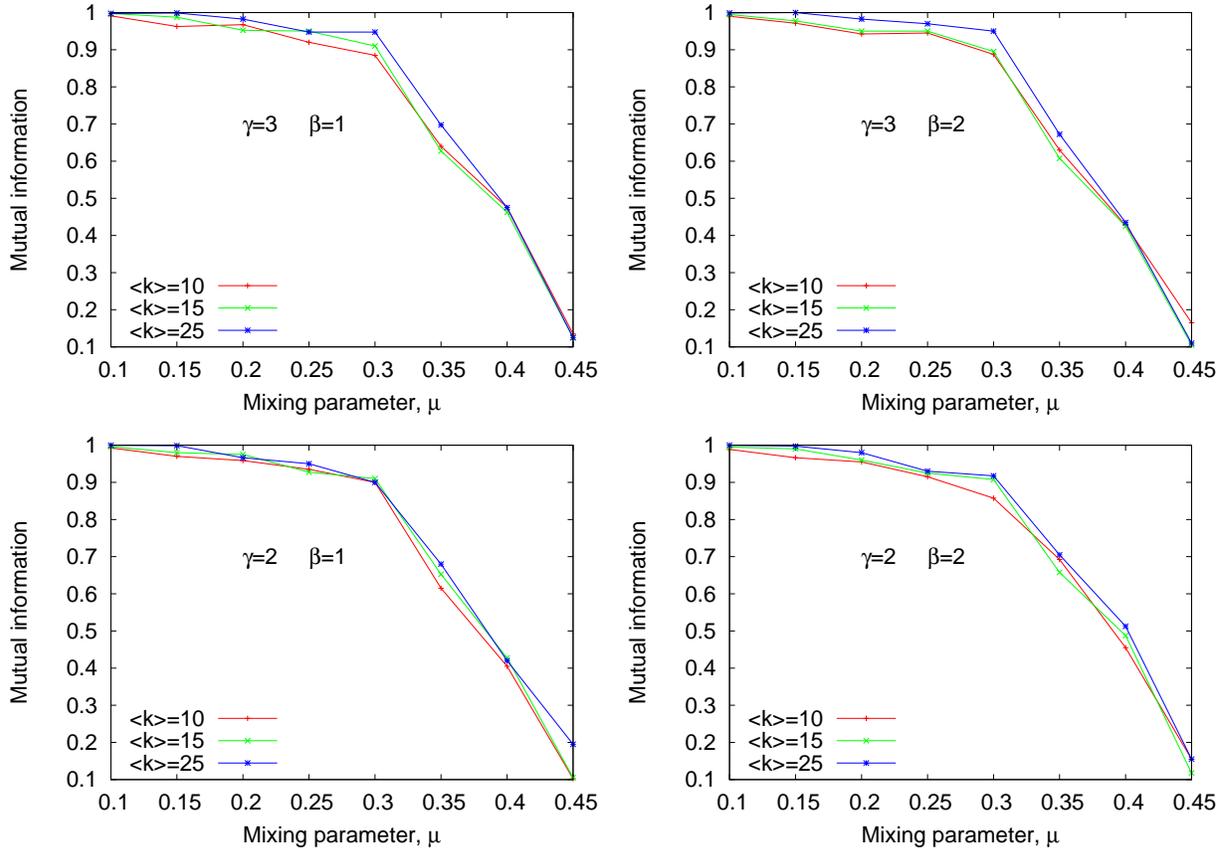}
\caption{Results of the benchmark of the random networks with N=1000. The data are presented when average node degree, $<k>$ is equal  10,15,25.}
\end{center}
\end{figure}

\begin{figure}[ht!]
\begin{center}
\includegraphics[angle=0,width=1\linewidth]{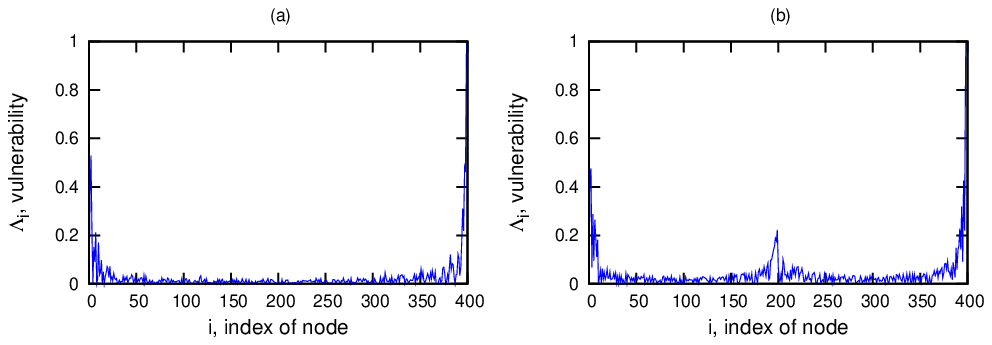}
\caption{In the figure (a) is shown the vulnerability of a network which has only one Erdos-Renyi cluster, in the figure (b) the network consists of two Erdos-Renyi clusters with a small bridge between them. $i$ is index of node in the acyclic sequence.}
\end{center}
\end{figure}


\begin{thebibliography}{10}

\bibitem{Milo2002Network}
R.~Milo, S.~Shen-Orr, S.~Itzkovitz, N.~Kashtan, D.~Chklovskii, and U.~Alon,
  ``{Network Motifs: Simple Building Blocks of Complex Networks}'', {\em
  Science}, vol.~298, pp.~824--827, 2002.

\bibitem{Lehmann2003Citation}
S.~Lehmann, B.~Lautrup, and A.~D. Jackson, ``{Citation networks in high energy
  physics}'', {\em Phys. Rev. E}, vol.~68, no.~2, p.~026113, 2003.

\bibitem{CRGS10}
B.~Corominas-Murtra, C.~Rodriguez-Caso, J.~Go\~{n}i, and R.~Sol\'{e},
  ``{Topological reversibility and causality in feed-forward networks}'', {\em
  New J. Phys.}, vol.~12, p.~113051, 2010.

\bibitem{Costa2007Characterization}
L.~da F.~Costa, F.~A. Rodrigues, G.~Travieso, and V.~P.~R. Boas, ``{Characterization of
  complex networks: A survey of measurements}'', {\em Adv. Phys.}, vol.~56,
  no.~1, pp.~167--242, 2007.

\bibitem{KN09}
B.~Karrer and M.~E.~J. Newman, ``{Random graph models for directed acyclic
  networks}'', {\em Phys. Rev. E}, vol.~80, p.~046110, 2009.

\bibitem{Karrer2009Random}
B.~Karrer and M.~E.~J. Newman, ``{Random Acyclic Networks}'', {\em Phys. Rev.
  Lett.}, vol.~102, p.~128701, 2009.

\bibitem{Krivov2006OneDimensional}
S.~V. Krivov and M.~Karplus, ``{One-Dimensional Free-Energy Profiles of Complex
  Systems: Progress Variables that Preserve the Barriers.}'', {\em J. Phys.
  Chem. B}, vol.~110, pp.~12689--12698, 2006.

\bibitem{girvan:community-structure-social-bio-networks}
M.~Girvan and M.~E.~J. Newman, ``{Community structure in social and biological
  networks}'', {\em Proc. Natl. Acad. Sci. USA}, vol.~99, pp.~7821--7826, 2002.

\bibitem{Radicchi2004Defining}
F.~Radicchi, C.~Castellano, F.~Cecconi, V.~Loreto, and D.~Parisi, ``{Defining
  and identifying communities in networks}'', {\em Proc. Natl. Acad. Sci. USA},
  vol.~101, pp.~2658--2663, 2004.

\bibitem{SR95}
A.~J. Seary and W.~D. Richards, ``{Partitioning Networks by Eigenvectors}'',
  {\em Proceedings of the International Conference on Social Networks}, vol.~1:
  Methodology, pp.~47--58, 1995.

\bibitem{Newman_PRE74}
M.~E.~J. Newman, ``{Finding community structure in networks using the
  eigenvectors of matrices}'', {\em Phys. Rev. E}, vol.~74, p.~036104, 2006.

\bibitem{GfellerD_MDnetworks_2007}
D.~Gfeller, P.~De Los Rios, A.~Caflisch and F.~Rao,  ``{Complex network analysis of free-energy landscapes}'', {\em Proc. Natl.
  Acad. Sci. USA}, vol.~104, pp.~1817--1822, 2007.


\bibitem{Newman2011}
M.~E.~J.~Newman, ``{Communities, modules and large-scale structure in networks}'', {\em Nature Phys.}, vol.~8, pp.~25--31, 2011


\bibitem{Fortunato2010rev}
S.~Fortunato, ``{Community detection in graphs}'', {\em Phys. Reports}, vol.~486, pp.~75--134, 2010.

\bibitem{Schaeffer2007}
S.~Schaeffer, ``{Graph Clustering},'', {\em Comp Sc. Rev.}, vol.~1, pp.~27--64, 2007.

\bibitem{Prada-Gracia}
D.~Prada-Gracia, J.~Gomez-Gardenes, P. Echenique and F. Falo, ``{Exploring the Free Energy Landscape: From Dynamics to Networks and Back}'', {\em Plos. Comput. Biol.}, vol.~5, pp.~e1000415, 2009.


\bibitem{Wu2004Finding}
F.~Wu and B.~A. Huberman, ``{Finding communities in linear time: a physics 
 approach}'', {\em EPJ B}, vol.~38, pp.~331--338, 2004.

\bibitem{Rao2004Protein}
F.~Rao and A.~Caflisch, ``{The Protein Folding Network}'', {\em J. Mol. Biol.},
  vol.~342, pp.~299--306, 2004.

\bibitem{ER59}
P.~Erd\H{o}s and A.~R\'{e}nyi, ``{On random graphs}'', {\em Publ
  Math-Debrecen}, vol.~6, pp.~290--297, 1959.

\bibitem{Watts1999Small}
D.~J. Watts, {\em {Small worlds: the dynamics of networks between order and
  randomness}}.
\newblock Princeton University Press, 1999.

\bibitem{BENCH}
A.~Lancichinetti and S.~Fortunato and F.~Radicchi, ``{Benchmark graphs for testing community detection algorithms}'', {\em Phys. Rev. E},
vol.~78,pp.~046110,  2008.

\bibitem{Fortunato_compar}
A.~Lancichinetti and S.~Fortunato, ``{Community detection algorithms: A comparative analysis}'', {\em Phys. Rev. E},
vol.~80,pp.~056117, 2008.


\bibitem{Blondel}
V.~D.~Blondel, J.-L.~Guillaume, R.~Lambiotte, and E.~Lefebvre,``{Fast unfolding of communities in large networks}'', {\em J. Stat. Mech.-theory E.},
vol.~2008,pp.~10008, 2008.

\bibitem{Palla}
G.~Palla, I.~ Der\'{e}nyi, I.~ Farkas, and T.~Vicsek, ``{Uncovering the overlapping community structure of complex networks in nature and society}'', {\em Nature},
vol.~345,pp.~814--818, 2005.

\bibitem{Donetti}
L.~Donetti and M.~A.~Munoz,``{Detecting network communities: a new systematic and efficient algorithm}'', {\em J. Stat. Mech.-theory E.},
vol.~2004,pp.~10012, 2004


\bibitem{Clauset}
A.~Clauset, M.~E.~J. Newman and C.~Moore, ``{Finding community structure in very large networks}'', {\em Phys. Rev. E},
vol.~70,pp.~066111, 2004.

\bibitem{Amaral}
R.~Guimera and N.~Amaral, ``{Functional cartography of complex metabolic networks}'', {\em Nature},
vol.~433, pp.~895--900, 2005

\bibitem{Rosvall2007}
M.~Rosvall and C.~T.~Bergstrom, ``{An information-theoretic framework for resolving community structure in complex networks}'', {\em Proc. Natl. Acad. Sci. USA},
vol.~104, pp.~7327--7331, 2007.


\bibitem{Rosvall2008}
M.~Rosvall and C.~T.~Bergstrom,``{Maps of random walks on complex networks reveal community structure}'', {\em Proc. Natl. Acad. Sci. USA},
vol.~105,pp.1118-1123, 2008.

\bibitem{Ronhovde}
P.~Ronhovde and Z.~Nussinov, ``{Multiresolution community detection for megascale networks by information-based replica correlations}'', {\em Phys. Rev. E},
vol.~80,pp.~016109, 2009.

\end{thebibliography}
\end{document}